\documentclass[11pt]{article}
\usepackage{geometry}                
\usepackage{graphicx}
\usepackage{amssymb}
\usepackage{amsmath}

\begin{document}

\begin{center}
{\Large Tangled Nature: A model of emergent structure and temporal mode among co-evolving agents.}\\
\vspace{0.5cm} 
\vspace{1cm} 
{Henrik Jeldtoft Jensen}\\
Centre for Complexity Science and Department of Mathematics, Imperial College London, South Kensington Campus, SW7 2AZ, UK\\
and\\
Institute of Innovative Research, Tokyo Institute of Technology, 4259, Nagatsuta-cho, Yokohama 226-8502, Japan 
\end{center}

\tableofcontents

\section{Abstract}
Understanding systems level behaviour of many interacting agents is challenging in various ways here we'll focus on the how the interaction between components can lead to hierarchical structures with different types of dynamics, or causations, at different levels. We use the Tangled Nature model to discuss the co-evolutionary aspects connecting the microscopic level of the individual to the macroscopic systems level. At the microscopic level the individual agent may undergo evolutionary changes due to “mutations of strategies”. The micro-dynamics always run at a constant rate. Nevertheless, the systems level dynamics exhibit a completely different type of intermittent abrupt dynamics where major upheavals keep throwing the system between meta-stable configurations. These dramatic transitions are described by a log-Poisson time statistics. The long time effect is a collectively adapted of the ecological network.

We discuss the ecological and macroevolutionary consequences of the adaptive dynamics and briefly describe work using the Tangled Nature framework to analyse problems in economics, sociology, innovation and sustainability

\section{Complexity Science, Tangled Nature and Evolution}
The Tangled Nature model is more a framework than a specific model. It was introduced as a way to approach biological evolution from the perspective of Complexity Science with a focus on the emergence of a slow macro-dynamics from the fast micro-dynamics.  One may ask why Complexity Science is needed to describe evolutionary processes. The answer is that evolving organisms - and typically any system of evolving agents - involve a degree of co-evolution. The selection pressure experienced by one agent will at least in part be determined by the properties of other co-existing agents. What we have in mind can be anecdotally described by thinking of the triangle of monkeys, bananas and lions. The banana loving monkey will certainly have a much easier time in an environment with plenty of banana plants and few lions than if the monkey has to brave life surrounded by many lions and few bananas. This interdependence between organisms, or agents, makes evolutionary dynamics a prototype example of a complex system.

Complexity Science focus on the collective systemic level properties resulting from, or emerging from, multitudes of interdependent processes. Beside of biological evolution other examples include the dynamics of our brain, economics, finance, cancer etc. Complexity Science tries to understand how new emergent level of processes arise as one go up through the various hierarchies of a system.  In the case of the brain one level consist of interacting neurones. At the neuronic level we find complicated electrochemical processes, at the level of the entire brain the collective dynamics zillions of neurones leads to the processes responsible for our thoughts and feelings. 

What are the Complexity Science features of evolution that we would expect a generic modelling framework to capture? Obviously the influence between organisms already mentioned above should be included. We would then like to understand how much of the macro-evolutionary phenomenology at the systemic level can be reproduced by a model of reproducing individuals forming the basic level. The systemic level phenomenology involves non-stationary log-time phenomena like intermittency, related to punctuated equilibrium, consisting of long periods of little extinction or creation interspersed by relative dramatic mass extinctions and following creation events\cite{TaNaTimeDep,Anderson04,Jensen_Arcaute_2010}.  Also at long time scales one observe non-stationary behaviour with gradually decreasing rate of extinction since the Cambrian explosion\cite{Newman99c}. Of course this trend is currently replaced by a dramatic mass-extinction probably triggered by human activity. At the short time level, we will like our modelling framework to produce, say, abundance distributions and species area relations of shapes observed in nature.

And have we first established a theoretical platform able to reproduce consistently, at least at a qualitative level, observed feature of evolutionary ecology, we can then use our modelling to address important theoretical question which are hard to settle by direct observation. These will be questions such as the role of group selection versus selection at the level of individuals\cite{Brinck_Jensen_TE_2018} or we may study some of the information theoretic measures suggested as being optimised during the evolution of an ecosystem\cite{Jones_2010A,Brinck_2017}.

After these comments it should be clear that the Tangled Nature model is intended to be both a model of evolution and a paradigmatic model of complex systems dynamics in general. But how is it possible that the dynamics of some simplistic agents simulated on a computer can capture the behaviour of the extremely complicated biological, or socio-economical, world surrounding us. The answer to this question is essential to understand the agenda of Complexity Science. As a long time ago argued by Alfred North Whitehead\cite{Whitehead_1919,Whitehead_1929} processes constitute the basic building blocks of the world. We may sometimes think the world consist of "things". Small hard atoms combining to molecules, combining to proteins, combining to organism and so on. But when you start to think about it, the atoms, the molecules, the proteins and the organisms are all processes co-evolving  and interdependent. Our world doesn't really consist of components like LEGO bricks but together to build structures. The hierarchies of structures we observe are hierarchies of intertwined processes. Why is this relevant to the agenda of Complexity Science and how we build models? Because when we think of the world as collections of processes, we realise that a model like Tangled Nature may very well deal with agents very different from the real agents of biology, or economics, though still include some relevant features of the processes making up a biological organism. For instance, some of the consequences of the fact that the success of the reproduction process is influenced by other coexisting organisms, can probably be captured, or summarised, by models emphasising on interaction between types. When this is the case the results may be of equal relevant to the understanding of certain properties of very different systems, say field ecology, tumour growth or economics. It is for example possible that the development of diversity among the agents involved exhibit similar dynamics in all these cases. 

So much for philosophical arguments for why models using simple components can be able to explain complicated emergent phenomena in complex systems. Now let us focus on the detail of the Tangled Nature model.

\section{The Tangled Nature Model}
Let us now present the mathematical structure of The Tangled Nature model. As in the paper that introduced the model\cite{TaNaBasic} we used TaNa as the shorthand for the model, but other authors have used other names and abbreviations, e.g. TNM, but since TNM is a standard classification system for cancer tumours it may cause less confusion if we stick to the label TaNa. 

The model is individual based and consist at time $t$ of $N(t)$ individuals, or agents, of various types labelled by ${\bf S}$. An agent can reproduce, with or without mutations and an agent can die. These are the only agent activities included in the basic version of the model, see e.g. \cite{TaNaTimeDep}. Precisely because we want to study how the temporal mode of the macro-level emerges from the micro-level dynamics, we let the agents generate all dynamical changes in the model. This is different from e.g. the  evolutionary dynamics introduced by Jain and Krishna \cite{Jain_Krishna_1998,Jain_Krishna_2002} in which a fast dynamics for the populations of species is defined in parallel with a slow dynamics for the interaction network between the species.

The offspring production occurs according to a type dependent reproduction rate $p_{off}({\bf S})$. During the reproduction event two copies of the mother are made and the mother is removed. Obviously  inspired by microbial reproduction. A mutation offspring ${\bf S}'$ will be of a type different from the mother ${\bf S}$. This is the mechanics which allows the population to move into hitherto unoccupied regions of type space. Let $n({\bf S},t)$ denote the number of agents of type ${\bf S}$ at time $t$.  This number may increase due to two effects. Reproduction of type ${\bf S}$ without any mutations will lead to $n({\bf S},t)\mapsto n({\bf S},t)+1$. If a type ${\bf S}'$  different from ${\bf S}$ reproduces and produces a mutant of type ${\bf S}$, we will also have $n({\bf S},t)\mapsto n({\bf S},t)+1$. The occupation of type ${\bf S}$ may decrease due to two effects. Firstly, if  both offspring produced by a reproduction event are mutants we'll have $n({\bf S},t)\mapsto n({\bf S},t)-1$. Secondly, if at time $t$ an agent of type ${\bf S}$ dies, we subtract one unit from $n({\bf S},t)$.

The definition of the reproduction probability is the most important feature of the model dynamics. The ability to reproduce summarises all the different kinds of influences the co-existing agents may have on each other. This is done through a weight function

\begin{equation}
H({\bf S},t) = \frac{k}{N(t)}\sum_{{\bf S}'} {\bf J}({\bf S},{\bf S}')n({\bf S}',t) -\mu N(t),
\label{ecology}
\end{equation}  
which will determine the offspring probability in the following way
\begin{equation}
p_{off}({\bf S},t) = \frac{\exp[H({\bf S},t)]}{1+\exp[H({\bf S},t)]}.
\end{equation}

Let us explain Eq. (\ref{ecology}). The matrix element ${\bf J}({\bf S},{\bf S}')$ describes the effect one agent of type ${\bf S}'$ has on an agent of type ${\bf S}$ and the efficiency with which this effect act is weighted by the relative importance $n({\bf S'},t)/N(t)$ of agents of types ${\bf S}'$ out of the total population. The constant $k$ simply determines the overall scale. The constant $\mu$ describes how all the agents $N(t)=\sum_{\bf S} n({\bf S},t)$ compete for shared resources and in this way corresponds to a kind of carrying capacity parameter. One may also think of $\mu$ as similar to the chemical potential in the grand canonical ensemble of statistical mechanics.  In principle one can imagine to obtain information about ${\bf J}({\bf S},{\bf S}')$ from population dynamics experiments where one investigate the effect on the reproduction rate of type ${\bf S}$ when ${\bf S}'$ is present. In practise this is of course not at all easy, since we have in mind all possible kinds of influences including trophic, mutualistic and antagonistic. In the case of bananas ${\bf S}_b$, monkeys ${\bf S}_m$ and lions ${\bf S}_l$ we will have $J({\bf S}_b,{\bf S}_m)<0$ and $J({\bf S}_m,{\bf S}_b)>0$ and similar for the ${\bf J}$ elements between monkeys and lions. Since lions don't care much for bananas it seems reasonable to assume  $J({\bf S}_l,{\bf S}_b)=0$ but it is possible of course that the nutrition provided by the lion manure should be captured by an element $J({\bf S}_b,{\bf S}_l)>0$.  

These considerations make it clear that it will be difficult to determine the entire matrix ${\bf J}({\bf S},{\bf S}')$ from data, not least because the matrix also has to contain the interaction strength between types that has not so far been brought into existence through the dynamics of mutations. Say e.g. the effect of a certain strain of bacteria that perhaps arrive next year due to mutations of the good old and ubiquitous  E. coli bacteria.  If one is able to study a limited number of types \cite{Rivett_Bell_2018} one may perhaps be able to estimate the elements of ${\bf J}$ between extant types, but even then it will be difficult to estimate the elements between extant types and future mutant types. To circumvent these problems we can assume the elements $J({\bf S},{\bf S}')$ to be random numbers and study how the dynamics of the model depend on the statistical properties of $J ({\bf S},{\bf S}')$.

In the simplest case one ignores correlations  between the matrix elements. This is of course not a realistic assumption since similar types (think of chimpanzees and orangoutangs) should be related to other possible types through set of similar interdependencies, i.e. we would like
\begin{equation}
 {\bf J}({\bf S_{chimp} },{\bf S}')\approx {\bf J}({\bf S_{orang} },{\bf S}')\; {\rm most} \; {\bf S}'.
\end{equation}

Correlations can be included in various ways, details are discussed in \cite{lair05:tang}, \cite{lair07:EcoMod} and a particular elegant method was  invented by Andersen and Sibani\cite{PhysRevE.93.052410}. 

The label ${\bf S}$ can be defined in many ways, for example as a string of binary attributes, or "genes", ${\bf S}=(S_1,S_2,...,S_L))$, with $S_i=\pm 1$. In this case the model describes the evolution of the occupancy on the corners of an $L$ dimensional hyper cube. Mutations consist of a gene swapping sign during the reproduction, so $S_i^{daughter}=-S_i^{mother}$ with probability $p_{mut}$. Hence one mutation will place an offspring on one of the adjacent corners of the hyper cube. I this way mutants perform a connected random walk through type space. 

An agent is removed with probability $p_{kill}$. In the simplest case this probability is independent of type and time. 

So to summarise: at each time step an individual is chosen at random with uniform probability $1/N(t)$. The individual is allowed to reproduce with probability $p_{off}({\bf S})$. Next another individual is again picked at random with probability $1/N(t)$ and this individual is removed with probability $p_{kill}$.  Recall the practise used for Monte Carlo (MC) simulations of, say, lattice models with $N$ sites, where a lattice site is picked at random with uniform probability $1/N$ and then the MC update attempted. Here one define a macroscopic time step as a "sweep" of the entire lattice, so $N$ update attempts. For the TaNa model we define the macroscopic time scale as $N(t)/p_{kill}$ reproduction and killing attempts. This is on average the number of  removal attempts needed to remove an population of size $N(t)$, so we call this a generation.   

Obviously, the details of the definitions above can be varied according to the demands of relevance in different situations. Examples will be given below. 

\section{Phenomenology}
We'll now sketch the structures and temporal behaviour resulting from the simple individual based dynamics. Our focus will be on the behaviour at the systemic level and how it differs from the level of the individual agents. Think of the ideal gas. It may consist of about $N=10^{23}$ particles, but since the particle do not interact the statistics of a single particle is identical to the statistics of the entire gas in the sense that pressure and temperature are simply given directly in terms of single particle averaged. The temperature is of course given in terms of the average kinetic energy of the individual molecules according to $\frac{3}{2}k_B T= \langle \frac{1}{2} mv^2\rangle$ and the pressure is $N$ times the momentum transfer per area of each particle. Interactions allow  new emergent behaviour to appear at the level of the entire system. Perhaps the most direct example of this in the TaNa model is the difference between the dynamics as it appears at the level of the individual agent and the dynamics of the configurations occupying type space. At the level of individuals we have agents appearing at random instances as a consequence of the random reproduction events. An agent stays in the system until it is removed through the random killing event. This is a Poisson process since the probability that an individual is removed from the system during one generation is the probability the individual is selected in one attempt, $1/N(t)$, times the number of attempts constituting constituting a generation, $N(t)$, times the probability the individual is removed  $p_{kill}$, i.e per generation all individuals may be removed with the constant\footnote{Note when defining a generation, we neglect the change to $N(t)$ during the the $N(t)$ update attempts, this approximation turns out not to cause any problems} probability $p_{kill}$. So we know that the distribution of life times will be exponential as it is for Poisson process.

Let us now turn to the systemic level of configurations in type space as described by the function $n({\bf S})$. Here the dynamics is very different. Figure 1. indicates the temporal evolution of the configurations. We notice that the dynamics is intermittent with periods where the occupancy doesn't change much separated by rapid transitions from one meta-stable configuration to another.

\begin{figure}[!h]
\centering
\includegraphics[width=11cm,height=6cm]{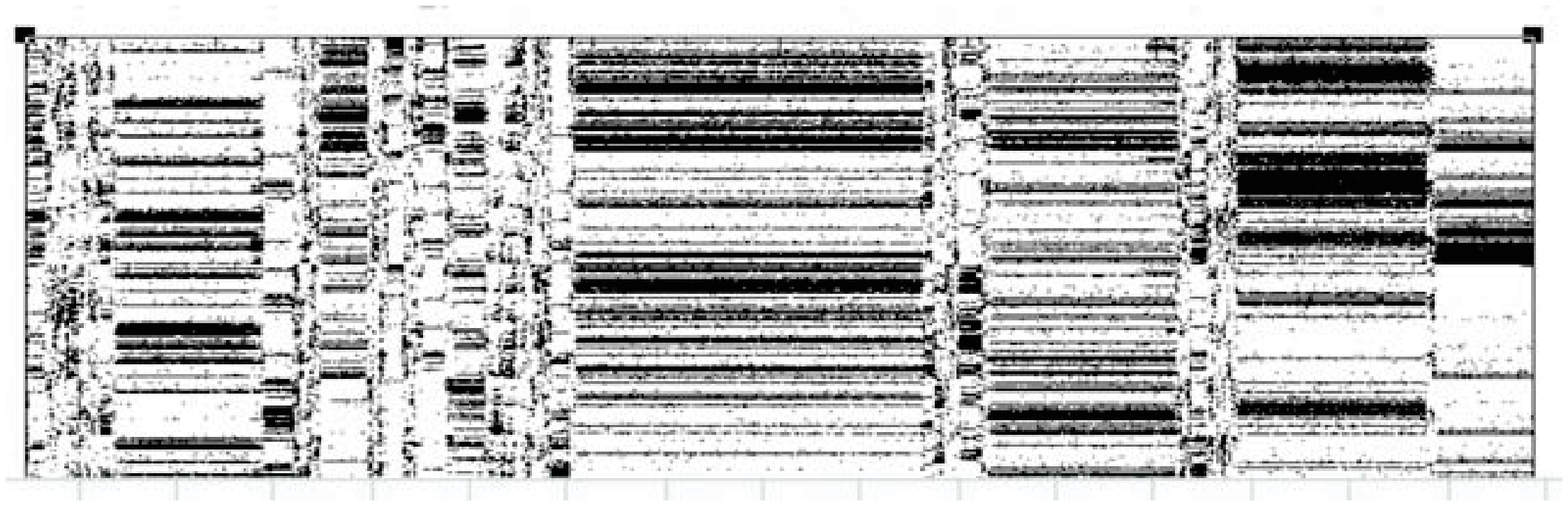}
\includegraphics[width=11cm,height=6cm]{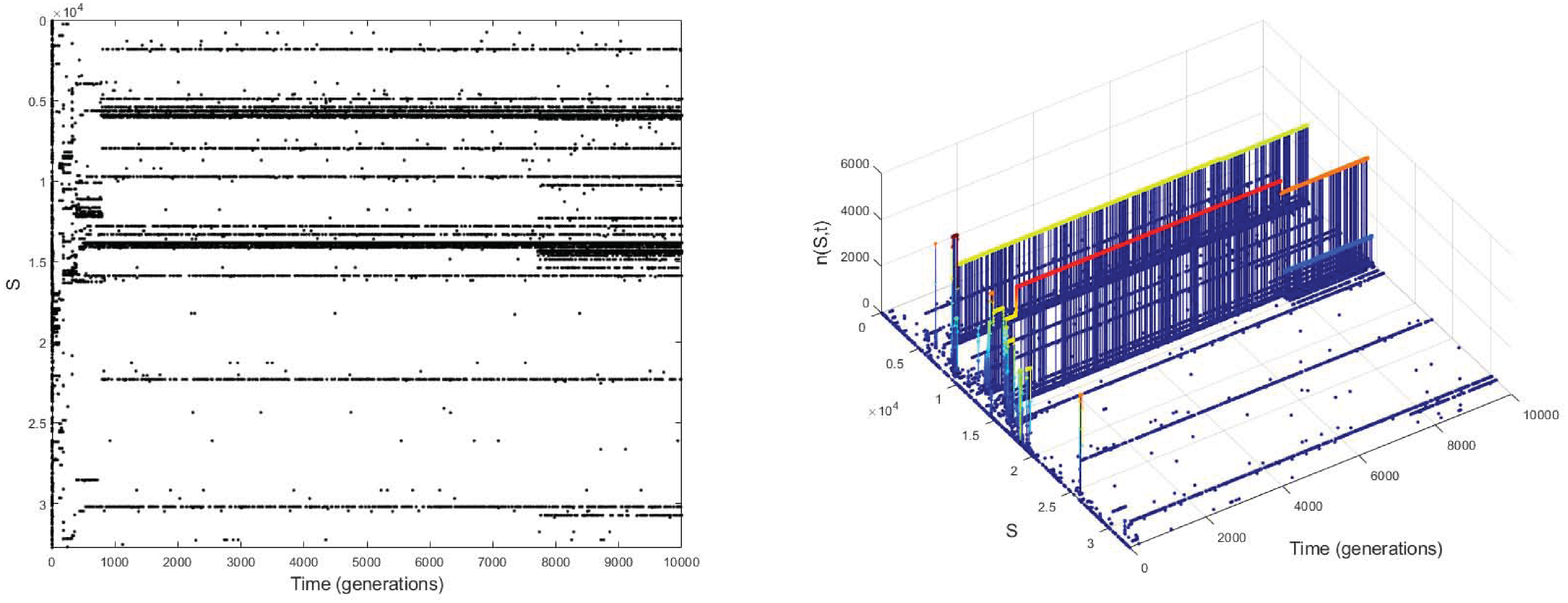}
\caption{Intermittent dynamics in type space. Time in generations is along the x-axis. In the top and the bottom left figure the y-axis is a number labelling the types running from 1 to $2^L$. If a type is occupied at a given time, a black dot is place at the y-value corresponding to the type. The z-value in the three dimensional plot bottom right shows the actual size of the population at time $t$ each type. One notice the q-ESS states in which a few well occupied types, the core of the ecology, are present continuously. The core is surrounded by a cloud of weakly occupied mutant types originating from reproduction events in the core. 
The parameters for the plots are as follows:  $\theta = 0.25$, $p_{mut} = 0.0125$, $p_{kill} = 0.2$, $k = 14.3$, $\mu = 0.005$. and $L=20$ for the top panel and $L=15$ for the bottom. The top plot is curtesy to Matt Hall and the bottom to Lorenzo Palmieri.
}
\label{Intermit}
\end{figure}

So although at the level of individuals everything happens at a steady pace, at the level of configurations, corresponding to, say, the taxonomic level of species, abrupt changes interrupt essentially static periods. This is similar to the notion of Punctuated Equilibrium \cite{Eldredge_Gould_1972} and is an example of how the dynamics in a complex system at the systemic level can be entirely different from the dynamics at the level of the components. In the TaNa model there are a number of emergent time scales. The life time of a species, which is the number of generations a line in Fig. \ref{Intermit} extends over. Next we have the duration of the meta-stable, or quasi Evolutionary Stable Strategies (q-ESS) \cite{TaNaBasic}, which corresponds to the number of generations of the more or less static sections of parallel lines in Fig. \ref{Intermit}  and the finally the duration of the transition between the q-ESS states. This transitions occur on a much faster time scale than the lifetime of the q-ESS and can therefore not be seen in Fig. \ref{Intermit}. All three timescales are distributed according to approximate power laws, see \cite{TaNaBasic}. 

It is a typical feature of complex systems to have vastly different kinds of dynamics at different scale. For illustration think of the brain. The individual neurones fire on the time scale of microseconds. The response time to external stimuli is of order 200 ms and it can of course take a very long time before the human brain finds the correct solution to a certain problem, e.g. a chess move. The separation of various time scale in the TaNa model are equivalent to what is found in real evolutionary-ecology in terms life times of individuals, of species and of entire ecosystems. 

The simple individual agent based reproduction, mutation and killing of the TaNa leads to a  range of behaviours which are also observed in real evolutionary systems. Let us here just briefly mention some of these with references to papers that contain more details. 

{\bf Non-stationary dynamics} The individual agent  is subject to an adaptive pressure from the surrounding co-existing configuration of agent types. This leads gradually to configurations that are better "tuned" in terms of their mutual influences. Let the distribution of couplings  ${\bf J}({\bf S}_1,{\bf S}_2)$ between any to randomly chosen types ${\bf S}_1$ and ${\bf S}_2$ be uncorrelated  and symmetric around zero.  One observe that the subset of couplings connecting types which are actually extant (i.e.  occupied) very slowly becomes biased towards positive, i.e. mutualistic, couplings\cite{TaNaNetwork,lair05:tang,Becker_Sibani_2014}, a result consistent with recent observations\cite{Rivett_Bell_2018}

The collective adaptation of the configurations in type space has various consequences in terms of how the properties of the system develops with time: Firstly, the q-ESS states become longer lived q-EES\cite{TaNaTimeDep}.  The mechanisms behind the increased configurational stability was studied in detail by Becker and Sibani\cite{Becker_Sibani_2014}.  They found that the q-ESS states consist of a network of relative few but highly occupied types, equivalent to the wild types in ecology. This set of types forms a relatively robust core for the ecosystem and is surrounded by a cloud of weakly occupied types produced when reproduction of the core types mutate and thereby move off-springs to the adjacent sites in type space.  The collective adaptation is also related to an increase in the correlations between the couplings of extant types time\cite{Jones_2010A,Jones_PRE_2010}. Even when the couplings across all possible types ${\bf S}$ are uncorrelated, the adaptive search resulting from the evolutionary dynamics selects sub sets of types $\{ {\bf S}_1,{\bf S}_2,...,{\bf S}_D\}$, which constitutes ever more well adapted configurations. The search sequence of on average better adapted q-ESS configurations in type space continues until some maximally stable, or at least practically ineradicable, configuration is found. The length of the transient dynamics towards this stable configuration grows exponentially with the length $L$ of the vector ${\bf S}, see $\cite{TaNaTimeDep}. For small values of $L$ e.g. $L=10$ the switching between q-ESS states stops after a number of generations of order $10^5$, but already for $L=20$ the switching doesn't settle for a number of generations that are numerically affordable. The transient during which the statistics of the TaNa is non-stationary is logarithmically slow\cite{Anderson04} in the sense that the average population size $N(t)$ and diversity $D(t)$ increases linearly with $\log(t)$.  The same logarithmic time dependence is often seen in physical systems such as relaxing glasses\cite{Sibani2013}.

\subsection{Record Dynamics - and ageing} The transitions between q-ESS states occurs at random times and follows a non-stationary statistics also found for record dynamics\cite{Anderson04,Sibani2013}. Record dynamics consists in drawing a number from an infinitely large uncorrelated set at each time step. Each time one manage to draw a number that is larger than any seen before, one has a record. When this happens, the time time step is called a record time and one can show \cite{Sibani2013,Nevzorov01} that these times to a good approximation are distributed as a Poisson Process in logarithmic time. Many non-stationary processes are found to follow this kind of dynamics see e.g.\cite{Richardson_2010,SIBANI201136,Sibani2013}. The hallmark of such dynamics is that the properties of the system change logarithmically slowly with time. This is not meant to say, that, e.g. the centre of mass of the system moves logarithmically slowly. It is the intrinsic systemic properties that change, e.g. its stability. For this reason one talk about ageing in the same sense as, say, a human ages with time. A 50 year human clear has some very different abilities -- properties -- than a one year old. 
The ageing aspects of the TaNa dynamics were studied by \cite{Jones_PRE_2010,Becker_Sibani_2014}.

\subsection{Abundance Distribution} The configurations found during the q-ESS states can be throughs of as equivalent to ecosystems and it is therefore interesting to compare the statistics of these configurations to observed statistics. The composition of an ecosystem is often described in terms of the Species Abundance Distribution (SAD), which very often is found to be well approximated by a log-normal distribution\cite{TaNaNetwork}.  In the TaNa model the SAD evolves towards such a distribution when the matrix ${\bf J}$ has a sufficiently large number of non-zero elements. Think of ${\bf J}$ as the adjacency matrix of a network with nodes at each the $2^L$ types ${\bf S}$. Let $\theta$ denote the fraction of non-zero elements $J({\bf S}_1,{\bf S}_2)$ and define two nodes ${\bf S}_1$ and ${\bf S}_2$ to be adjacent, i.e. connected by a link,  if a non-zero coupling exists between the two types. 
From the theory of random networks consisting of  $\Omega$ nodes,  it is known that for $\theta>1/\Omega$ an "infinite cluster" exists\cite{Newman_Ntw_book}. By that we mean that a large fraction proportional to $\Omega$ belongs to the same cluster in the network of all $\Omega$ nodes. In the version of the TaNa model that uses the hypercube as its type space $\Omega = 2^L$ where $L$ is the length of the genome. It was found in \cite{TaNaNetwork} that the SAD takes the lognormal functional shape when $\theta>2^{-L}$. This suggests hat the observed log-normal species abundance distribution may indicate that all living beings belonging to the same interaction network.

\subsection{Species Area Relation} The spatial distribution of species in an ecosystem is characterised by the Species Area Relationship (SAR)\cite{laws05:spec}. A power law relation between number of species $N$ within an area of size $A$ in the form
\begin{equation}
N\propto A^z,	
\label{SAR_eq}
\end{equation}
has been observed very broadly in ecology and has accordingly attracted lots of attention. The exponent  $z$ depends or the type of species considered and the geography of the location studied. In fact the power law relationship is not always found\cite{SpecDiv}, but for non-island geographies it is frequently encountered.   A similar relation is found in the TaNa when one place a version of the model on each lattice point of  e.g. a square lattice. The types are allowed to move about by performing a random walk from one lattice site to another. The SAR relation in Eq. (\ref{SAR_eq}) comes about as a consequence of two opposite directed dynamical trends. Firstly mutations will locally created new extant types and this leads to heterogeneity between different regions of the lattice. On the oner hand  and diffusion will mix the population from different regions and in doing that tend to make the population across the lattice homogeneous. For these reasons, the exponent $z$ depends on the mutation rate and the amount of diffusion. This dependence may correspond to dependencies observed in real data. In the TaNa model one for example finds that the value of the exponent $z$ decreases with increasing diffusion rate (types moving more easily around from site to site) which agrees with the observation that $z$ for birds is typically lower that the value for land species\cite{laws05:spec}

\subsection{Entropy and emergence of ecological structure} Ecologist have proposed information theoretic measures that the evolution of ecosystems are expected to increase. 
 
 Ulanowicz considered\cite{Ulanowicz_1986,Ulanowicz_2011} the flow $T_{ij}$ from species $i$ to species $j$ and defined a set of measures inspired by information theoretic concepts. An entropic measure of the flow (of energy, material, etc.) is defined as
 \begin{equation}
 	H = -\sum_{i,j}\frac{T_{ij}}{T{..}}\log\frac{T_{ij}}{T_{..}}.
 \end{equation}  
The Capacity is defined as 
   \begin{equation}
 	C = -\sum_{i,j}T_{ij}\log\frac{T_{ij}}{T_{..}}.
 \end{equation}  
The Mutual Information is defined as  
 \begin{equation}
 	MI = \sum_{i,j}\frac{T_{ij}}{T{..}}\log
	\frac{\frac{T_{ij}}{T_{..}}}  {\frac{T_{i.}}{T_{..}}\frac{T_{.j}}{T_{..}}}.
 \end{equation}  
The notation is as follows
\begin{eqnarray}
T_{i.}&=\sum_k T_{ik}\nonumber\\
T_{.j}&=\sum_l T_{li}\\
T_{..}&=\sum{ij} T_{ij}\nonumber.
\end{eqnarray}
The link to  probability or information theoretic definitions is established by thinking of a total number of flow units given by $T_{..}$. If the units are distributed uniformly random on all the links, the probability to find a specific unit on link between $i$ and $j$ is simply $T_{ij}/T_{..}$. 

The hypothesis that the evolution of an ecological network will be related to the maximisation of a certain measure is probably inspired by the 2nd law of thermodynamics, which states that the entropy of a closed system cannot decrease. Hence, only those temporal scenarios that corresponds to the entropy not decreasing can occur in a closed system. Similarly for systems of coevolving agents, we would like to have a measure that can distinguish possible trains of events from impossible. The measures proposed by ecologist typically involves the energy flow between the species of an ecosystem and are therefore very difficult to test on real systems over evolutionary time scales. The TaNa model offers an opportunity to test the suggested measures within a theoretical framework that has demonstrated at least qualitatively to reproduce many features observed on real systems. A recent study \cite{Brinck_2017} of a trophic version of the TaNa model found that the long time effect of the evolutionary dynamics leads to a relative ascendancy
 \begin{equation}
  	\frac{A}{C}=\frac{MI}{H}
 \end{equation}
  around 0.4 and maximises the measure 
 \begin{equation}
 F=-e\frac{MI}{H}\ln\frac{MI}{H}.	
 \end{equation}
Ulanowicz\cite{Ulanowicz_1986,Ulanowicz_2014} has suggested that this corresponds to systems developing organisation while also remain sufficiently flexible to be able to continue to change and adapt. This is the ecosystems equivalent to Wagner's \cite{Wagner_2005} considerations at the level of individuals concerning adaptability and evolvability.  

The temporal evolution of two different entropic measures in the TaNa model was explored by Roach et al. \cite{RNSRFS_2017}. They define one entropy related to the SAD and another that directly relates to structure in type space of a few core types surrounded by a cloud of mutant types\cite{Becker_Sibani_2014}. The stochastic adaptive dynamics leads to a logarithmically slow increase with time in entropic measures of the entire population, this is related to an increase in diversity. In contrast the entropy per individual decreases with time, this is associated with emergent structures in type space. That the model simultaneously  exhibit two types of dynamics was also observed in a study of how adaptation makes the extant types become more correlated\cite{Jones_2010A}. It was found that the mutual information, which is a measure of correlation, between couplings within the strongly occupied core types increases logarithmically with time, while at the same time the mutual information between couplings involving core and cloud or cloud to cloud couplings decreases meaning that while the core as a result of adaptation becomes a more correlated structure, the entire configuration, with surrounding cloud types, decorrelates as a result of the random mutations. 

The intermittency, or punctuated equilibrium, of the TaNa model was studied by Wosniack, da Luz and Schulman from the perspective of thermodynamics\cite{WOSNIACK2017113}. The authors define an energy, and entropy and a free energy and study how these behave differently in the q-ESS states and the hectic states separating the q-ESS sates.

\subsection{Error threshold} Obviously, mutations moves the population around in type space and there by "explore" configurations which consists of types interacting in different ways. Some configurations will have better sets of $J{\bf S}_1,{\bf S}_2)$ couplings that combine to more efficiently overcome the negative effect of the term $-\mu N(t)$  representing the competition for resources. When mutations occur they may accidentally hit types that were not occupied previously and which happens to be coupled to the existing population through a  set of favourable  couplings. Through this mechanics mutations at a low rate can lead to configurations that collectively adapt. We want to focus on two aspects: the error threshold and breaking of time reversal symmetry. 

The error threshold is related to the fact that if the mutation rate is too large, adaptation doesn't happen. This is because for high mutation rates the configuration surrounding a type ${\bf S}$ keeps changing frequently and the effect of a beneficial mutation, which is an increased in the offspring rate $P_{off}({\bf S},t)$ does not last long enough to have an effect on the size of the population of type ${\bf S}$. One says the beneficial mutations cannot lead to fixation in the population. So for high levels of mutation the natural selection doesn't result in adaptation this corresponds to the error threshold introduced by Eigen and collaborators\cite{Eigen_1998,Eigen2013} and was for the TaNa model explored in \cite{TaNaQuasi}. In a diagram with $1/k$ out along the x-axis and $p_{mut}$ along the y-axis, a broad region located at small values of $1/k$ and $p_{mut}$ exists in which the q-ESS dynamics is observed. So a region where the TaNa ecology is able to establish configurations of co-existing types that for a while exhibit robustness against the stress caused by the $-\mu N$ term and the bombardment of new mutants.  Outside this region at low coupling and high mutation rate, adaptation doesn't happen. The two regimes are separated by a curve indicating the error threshold value of $p_{mut}$ as function of the over all strength of the couplings $J({\bf S}_1,{\bf S}_2)$.

\subsection{Breaking time reversal symmetry} The second interesting aspect of collective adaptation concerns the breaking of time symmetry in the following sense. At the microscopic level mutations at random switch the sign of one or more of the coordinates of the "genome" ${\bf S}$ of the mother to produce a genome ${\bf S}'\neq{\bf S}$ for the offspring. Mutations are equally likely to be beneficial as to be detrimental. By beneficial we mean that the change $\delta H=H({\bf S}')-H({\bf S})>0$ leading to $p_{off}({\bf S}')>p_{off}({\bf S})$. For detrimental mutations we have the opposite effect, $\delta H<0$. If the probability for these two cases are equal, how can it then be that better adapted configurations are produced over long times. This is related to the convex shape of the offspring probability. To have on average an approximate balance between reproduction and killing, the system moves to the regime where $p_{off}({\bf S})\simeq p_{kill}$. For values of $p_{kill}\ll 1$ where metastable q-ESS configurations can be established we find $p_{off}$ to be concave. Which means that the increase in $p_{off}({\bf S}_+)$ for a mutant ${\bf S}_+$ that has a weight function  $H({\bf S}_+)=H({\bf S})+|\delta H|$ is numerically larger than numerical value of the decrease suffered by a mutant which has a weight function $H({\bf S}_-)=H({\bf S})-|\delta H|$.  This means that the increase in the reproduction rate of the offspring resulting from the beneficial mutation is larger than the decrease of the reproduction rate of the detrimental mutation even if the change to the weight functions $H\pm|\delta H|$ occurs with equal probability. A test of this explanation is given in Sec. 12.6.5 in \cite{Sibani2013} by considering a piece wise linear functional shape for $p_{off}(H)$, which accordingly has now convex parts. When no convex part is present, long time adaptation of more stable q-ESS is not observed in the simulations. Is it plausible that biological evolution towards better adapted communities relies on the mathematical detailed shape of the offspring probability? Yes, it could  be the case since any function that goes from zero at minus infinity to 1 at plus infinity, as is the case for our offspring probability as function of the weight $H$, will have convex parts if it is a smooth differentiable function. What we just have noticed for the TaNa model can be seen as a mathematical explanation of how neutrality between beneficial and detrimental mutations can be compatible with a net positive effect of the beneficial mutations. So a mathematical suggestion for how   Darwin's explanation can work: It is well know that Darwin in \textit{The Origin of Species} explains  that because of the perpetual competition, organisms with beneficial mutations will be the ones that are able to dominate in future generations.

\subsection{Selection level} For may years it has been discussed at which level selection acts. This discussion can be paraphrased as individual selection versus group selection\cite{Wilson_1974}. The discussion is about whether the selection only happens at the level of the reproducing individual, or if  individuals can form groups which selection then acts upon. Intuitively the situation perhaps seems quite straight forward and one might think that it is fairly obvious that formation of social structure such as packs of wolfs or tribes of humans would evidently add to the ability to compete with other species. Or one might observe that the reproducing unit, say a human being, is in fact not just a unit but involves for example an entire ecosystem as the gut flora. Nevertheless, the discussion continues since so far it has been difficult to quantify the amount of selection pressure, or adaptation, that resides at group level versus the level of the reproducing individual. 

The TaNa model does make it possible, at least to some extend, to address the relative importance of the two kinds of selection in a quantifiable way. If an individual is selected on account of the membership of a certain group there most exist some dependency between the group and the individual. If selection only act at the level of the individual with the collective group level simply trailing as the linear sum of the adaptation of the individuals, we have that information only flows from the individuals to the group, or we will only observe bottom-up relationships. If, on the other hand, adaptive structures are formed at the collective we should see information flow from the top towards the bottom. The relationship between the individual and the group may perhaps be thought of as similar to the relationship between the genetic level and the phenotypical level. The phenotype of an organism is influenced by the genome, but the chance that the genome is carried on in the population is determined by the viability of the phenotype.  

In the study of the TaNa model the information flow was quantified by use of the Transfer Entropy\cite{Schreiber_2000} 
\begin{equation}
TE_{X \rightarrow Y} = \sum P(Y_{n+1},Y_n,X_n) \ln \frac{P(Y_{n+1}|Y_n,X_n)}{P(Y_{n+1}|Y_n)}.
\end{equation}
between two time series $X(t)$ and $Y(t)$. One time series $X(t)=n({\bf S},t)$ characterise the level of the individuals, the other time series $Y(t)=N(t)-n({\bf S},t)$ describes the totality of all the other species. Bottom-up flow occurs if $TE_{m\rightarrow M}>0$ and top-down if $TE_{M\rightarrow m}>0$. It was found that early q-ESS states predominately show bottom-up information flow. As successive q-ESS states are established and vanish the top-down flow becomes more pronounced\cite{Brinck_Jensen_TE_2018}.

\section{The broader picture}
The  Tangled Nature framework has been applied to a number of biological and socio-economical complex systems where co-evolution is important. 

\subsection{1/f fluctuations in the fossil record}
One of the first was the study by Rikvold and Zia \cite{Rikvold_Zia_2003,zia03:fluc} used the TaNa framework to address the question concerning the long time fluctuations in the fossil record. To improve the computational efficiency they used a slightly different version of the TaNa model in with parallel update enabling them to study the very long time fluctuations in the diversity and total population size. The study was in spired by suggestions that the fossil record is consistent with so call $1/f$ fluctuations \cite{SoleFossRec} (see e.g. \cite{Jensen_1998} for a discussion of $1/f$).  These fluctuations are interesting because they corresponds to very long "memory" in the time signal, or more correctly temporal correlations that decay logarithmically slowly. Whether the fossil record actually does exhibit these very long correlations were questioned by Kirchner and Weil \cite{Kirchner_Weil_1998}  on the basis that the interpolation procedure used in \cite{SoleFossRec} could lead to such correlation in a time record with no correlations. Nevertheless, Rikvold and Zia established that extinction and creation in the TaNa does produce long time $1/f$ correlations.

\subsection{Food webs, correlations, and migration versus mutation.}
Rikvold has together with collaborators investigated a number of other complex systems aspects of the TaNa framework. Food webs and constraints\cite{Rikvold_2007,Rikvold_2007a,Rikvold_Sevim_2007}, the effect of gradients in the environment and positive mutualistic interactions between agents\cite{FGPR_2010} and the effect of correlations in the set of interactions between the individual types\cite{Sevim_Rikvold_2006,Sevim_Rikvold_2005}. The latter work is similar to work by Laird in \cite{lair05:tang,lair07:EcoMod}. Both studies found that a correlated interactions only make a minor difference to the overall behaviour of the model. More recently Andersen and Sibani\cite{PhysRevE.93.052410} introduced correlations in to the J-matrix in a away that makes it possible to started with the uncorrelated matrix and then gradually make the set of couplings seen by types separated by a certain Hamming distance more similar or correlated. The conclusion is again that correlations doesn't qualitatively alter the overall dynamics but Anderson and Sibani points out that a correlated J-matrix does make the model more realistic in the sense that families of significantly related types develop more readily. 

Real biological ecosystems are subject to the arrival of new types both as a result of mutations of existing types and by new types migrating into the geographical region from elsewhere. The arrival of migrating types from somewhere else and the appearance of a new type due to a mutation will both have an effect on the existing network of types.  Murase, Shimada, Ito and Rikvold used the TaNa framework to study similarities and differences in the challenge to ecosystem stability caused by migrants and mutants\cite{Murase_Shimada_Ito_Rikvold_2010}. Adding migration to the assembly of the ecosystem does make a big difference. The intermittent nature of the isolated system where only mutants can add new types is replaced by a smooth gradual random walk in genotype space and as a consequence the $1/f$ fluctuations found by Rikvold and Zia\cite{Rikvold_Zia_2003,zia03:fluc} now behave as $1/f^2$ which are typical of a signal that behaves like a random walk.

\subsection{ Tangled Economy} The broad brush general character of the TaNa model suggests that it can be modified to capture dynamics of other coevolving systems such as companies related to each other through links of transactions and other influences. A modified Tangled Economy Model version of the correlated TaNa model introduced by Simon Laird\cite{lair05:tang,lair07:EcoMod} was considered by Robliano\cite{Robalino_2012} form the perspective of of a network of companies. The vector ${\bf S}$ is now thought of as specifying the characteristics of a company, say the type (such as financial institution or construction firm etc.), location, etc., so all the {\textit traits} of the company and denote it by ${\bf T}^\alpha$ for company number $\alpha$. The function $n({\bf S},t)$, which in the ecological interpretation of the TaNa model describes the number of individuals of type ${\bf S}$ at time $t$, is now thought of as the "capital" of the company, i.e. some measure of the economical strength of the company and for this reason we replace $n({\bf S},t)$ by $C^\alpha(t)$, the capital of company number $\alpha$. Co-evolutionary dynamics now is now apply to the variable $C^\alpha(t)$. Of course companies cannot reproduce and mutate but the can innovate and start up new companies by investments. And companies do not die, but the can go bankrupt. Similar to how the reproduction probability for the ecological version of the model depends on the configuration of coexisting types, a company's potential for an increase or a decrease of its capital $C^\alpha(t)$ is taken to be determined through interactions with other coexisting companies in the following way.

Influence of other companies $\beta$ on a given company $\alpha$ depends on how similar their "trait" vector ${\bf T}^\alpha$ and ${\bf T}^\beta$ are andis represented by
\begin{equation}
H(\alpha, t)=a_1\frac{\sum_{\beta=1}^{N(t)}J(\alpha, \beta)}{\sum_{\beta=1}^{N(t)}C(\alpha, \beta)}-a_2\sum_{\beta=1}^{N(t)}C(\alpha, \beta)-a_3\frac{N(t)}{R(t)}.
\label{weightfct}
\end{equation} 
where the elements of the matrix $C(\alpha, \beta)$ are exponentially correlated as function of the distance  ${\bf T}^\alpha- {\bf T}^\beta$.
The function $R(t)$ describes the to amount of "resources" available. Each company utilizes one unit of resource while the company is alive or active. Let $N(t)$ denote the total number of companies, then $N(t)+R(t)$ is constant in time. 

The economical growth of company $\alpha$ is determined by the probability to {\em gain}:
\begin{equation}
P_{gain}=\frac{\exp[H(\alpha, t)]}{1+\exp[H(\alpha, t)]}
\end{equation}
Upon success, company $\alpha$'s capital increases  proportionally to its current capital $C^\alpha(t)$,  so that the gains (or losses) are proportional to companies size
\begin{equation}
C^\alpha(t+1)=C^\alpha(t)\left(1+c_g\frac{J^+(\alpha)}{J^{Tot}(\alpha)}\right)
\label{Cgain}
\end{equation}
where $J^+(\alpha)$ is the total strength of positive interactions, $J^{Tot}(\alpha)$ is the total sum of the absolute value of all of $\alpha$'s interactions, and $c_g$ is simply a gain control coefficient to ensure realistic growth.
Similarly, if unsuccessful, company $\alpha$ loses capital according to
\begin{equation}
C^\alpha(t+1)=C^\alpha(t)\left(1-c_l\frac{J^-(\alpha)}{J^{Tot}(\alpha)}\right).
\label{Closs}
\end{equation}
If $C^\alpha(t)$ goes below a certain threshold the company is considered as bankrupt and is removed fro the systems. On the other hand if $C^\alpha(t)$ increases above another threshold the company will invest by creating a new company in theneighbourhood of ${\bf T}^\alpha$. For more details see\cite{Robalino_2012}. 

This very schematic dynamics produces a number of what appear at least anecdotally to be realistic tendencies and statistics: trends towards monopolisation, lifetime distributions of companies, GDP growth and company size distribution. 

Thurner, Klimek and Hanel have introduce a model that quantitatively capture the evolutionary ideas concerning economics put forward by Schumpeter\cite{Thurner_Schumpeter_2010}. Their model consider the co-evolutionary dynamics of goods. Production creates new goods by combining existing goods and different goods influence the amount of other goods thorough competition and destruction. This formalism is in particular well suited to study  the Schumpeterian idea of destructive innovation, where the arrival of a new innovation make an existing good reductant (say cars and horse carriers)  is    being combined to produce other goods. This economics model share a great deal of similarity with the TaNa framework and the occupancy in "goods space" exhibit in particular intermittent dynamics like the one shown in Fig. \ref{Intermit}.

\subsection{Tangled Sustainability} How to find sustainable paths for the development of socio-economic structures like cities, regions, countries all the way up to the entire planet is increasingly acknowledge as an urgent challenge and attracts the attention of ever wider circles of people including the United Nations who has introduce the notion of Sustainability Indicators\cite{UN_2007}.  The indicators are of environmental, economic, social, institutional nature and are expected to influence each other. Vazquez et al. presents in \cite{Vazquez_Rio_CMJ_2015} an attempt to define and study an Entangled Sustainability Model. The model is similar to the Tangled Economy Model just discussed above. The model suggests a dynamics for the time evolution of the sustainability indicators where the increase or decrease of the strength of the indicators happens through a dynamics similar to Eqns. (\ref{Cgain}) and (\ref{Closs}). The strength of the interaction matrix elements $J(\alpha,\beta)$ were assigned in collaboration with sustainability experts.  The intermittent dynamics seen in the original TaNa model is also present in the Tangled Sustainability version. Sets of indicators possessing more than 80\% of the total indicator strength are identified and called Pareto sets. These sets turn out to be long lived and are therefore considered as sustainable combinations of the indicators.

\subsection{Organisational and cultural evolution}
Arthur, Nicholson, Sibani and Christensen (ANSC) have studied organisational evolution by reinterpreting the meaning of the agents and their interrelationships\cite{ANAS_2017}. They interpret the q-ESS intermittent dynamics with successive transitions caused by the arrival of new types (new agents with new strategies in this interpretation) as Schumpeterain shocks, this is related to the discussion in \cite{Thurner_Schumpeter_2010}. 

Nicholson and Sibani further developed the TaNa framework to make contact to cultural evolution\cite{Nicholson_Sibani_2016,ANAS_2017}. They consider a version which combines the reproductive dynamics of the original TaNa model with elements similar to Axelrod's \cite{Axelrod_1997} model of  cultural dissemination. The agents inherit the genome, up to possible mutations, from their mother, but in addition agents are also able to exchange strategies (also represented by sequences) like in the Axelrod models. This more elaborate model still exhibit intermittency, log-time dependence and populations that can be classified  as core and cloud\cite{Becker_Sibani_2014}.

\subsection{Forecasting} 
The often encountered sudden abrupt changes in complex systems  are sometimes called tipping points, crashes, transitions, quakes or similar. As we have seen above the TaNa models also exhibit  intermittent dynamics as is clearly seen in Fig. \ref{Intermit}. This raises the question if it is possible to forecast or predict such changes. We know from experience that predicting financial crashes, onset of epileptic seizure or earthquakes is very difficult. But people have tried in recent years to find methods that may work in certain cases. One prominent example is the suggestion by Scheffer and collaborators\cite{Scheffer_2012}. They argue that just before a transition fluctuations will become much larger and this increase in fluctuations can be used as a precursor and warning sign. This method may possibly work if one deals with situations where one, or a few, macroscopic quantities, say a concentration of a chemical or density of a vegetation,  are able to capture the dynamics of the specific many component complex system under consideration. But in some cases, an average parameter may not suffice. The situation in evolutionary dynamics, is that the dynamics generates new entities (a new virus - think of SARS or a new company - think of Google or amazon) and these new agents may sometimes have a crucial effect on the stability of the system. In such cases averaging, like in the Scheffer approach, over the existing system will not capture the effect of the new agents. One needs to monitor the effect on the stability of the system of the new agents created through mutations or innovation. The new agents will appear from the part of type space that Kauffman\cite{Kauffman2000} very descriptively  calls the adjacent possible. In references \cite{ForecastPRL_2014,ForecastJPA_2016} the authors describe how the transitions in the TaNa model can be predicted a few generations before they happen despite of the very high dimensionality and fully stochastic nature of the dynamics. The method relies on a  mean field approximation of the high dimensional dynamics describing the time dependence of $n({\bf S},t$. A stability analysis is carried out about the observed metastable q-ESS configuration. It turns out that the fluctuations in $n({\bf S},t)$ or the total population $N(t)$ cannot be used as precursors, but an 80\% forecasting success can be obtained by monitoring the time evolution of the overlap between the present configuration vector ${\bf n}({\bf S},t)$ and the eigenvector corresponding to the most unstable direction of the high dimensional stability matrix.   

To make the methodology practically applicable further development is necessary since in many cases one will not have access to the mean field equations describing the dynamics of the considered system. The analysis suggests that one may to some degree be able to forecast approaching transitions by focusing attention on weakly occupied new types that suddenly undergo rapid growth. When such new agents arrive they are frequently a sign of an approaching systemic reconfiguration. 

Kuehn et al. \cite{Kuehn_2015} have developed a forecasting procedure similar to the one just described. The authors consider what they call saddle-escape transitions where an external perturbation is assumed to push the system along an unstable direction at a high dimensional saddle point. The difference to the TaNa scenario is that no new entity is create, the transition occurs as a change in the configuration of a give set of variable, whereas in the TaNa the evolutionary dynamics produces new agents and it is  the effect of those the forecasting try to capture.

\subsection{Gaia}
In a recent innovative paper Arthur and Nicholson\cite{Arthur_Nicholson_2017} developed the TaNa model to address in a quantitative way evolutionary aspects of Loveluck's Gaia hypothesis\cite{Loveluck_2016}. Loveluck observed that the processes of life are themselves involved in the regulation of the conditions under which life exists. Arthur and Nicholson let the carrying capacity parameter $\mu$ depends on type and can in this way construct a model that allows one to distinguish between non-biotic and biotic contributions to the carrying capacity for a single type. The evolutionary dynamics of the model can then be used to study how adaptation is able to let an ecosystem develop the biotic contribution to improve their capacity to support life in the sense that the total population  $N$ increases while the frequency of the major extinction events decreases (less frequent q-ESS transitions).

\subsection{Bacterial resistance}
Lawrence Schulman adapted the correlated TaNa model in order to study the evolutionary aspects of bacterial resistance to antibiotics\cite{Schulman2017}. In this version of the model, it is assume that a bacterium with a certain subsequences, say $S_3=1, S_4=0, S_5=1$\footnote{Schulman assume the allele values to be $S_i=1$ or $S_i=0$ instead of the original used $S_i=\pm1$}  of the genome ${\bf S}=(S-1,S_2,...,S_L)$  is susceptible to the presence of an antibiotic. This is modelled by introducing a smaller  survival probability for the agents, here the bacterium, with the susceptible sequence. The bacterium can escape the effect of the antibiotics in two ways. Either by a mutation that changes the configuration of the otherwise susceptible sub-sequence, or by acquiring a certain {\em protective} subsequence configuration somewhere along its genome. This sub-sequence is assume to counteract the effect the antibiotics has on the susceptible sub-sequence. Schulman's version of TaNa differs in an important way from the original version by assuming that horizontal gene-transfer is possible during reproduction. This assumption obviously incorporates a realistic aspect of bacterial reproduction. Perhaps the most interesting result of the study is the observation that the occupancy of specific genome configuration fluctuates very strongly. It is found that the probability to find a fraction $\omega$ of individual bacteria with the specific sub-sequence decays as $\omega^{-a}$ with $a$ roughly around $1.5$. Hence the first moment diverges and one can take this to indicate that when a bacterial infection is treated with antibiotics the adaptive evolutionary dynamics of the bacteria in they response to the selection pressure from the antibiotics leads to very different evolutionary trajectories. This translates into the possibility of very different responses from patient to patient.

\subsection{Evolving networks}
The TaNa sees the interaction structures as given once and for all and keeps the matrix $J({\bf S}_1,{\bf S}_2)$ fixed over time, while the population moves around n the type space of agents ${\bf S}_\alpha$. The properties of the types of the TaNa model are essentially defined through their interaction with other types and therefore if a type change through mutation it moves to an other position in type space and will then be subject to the interactions wired to that position. This is of course to some extend a matter of convention. One may also think of the interactions between types to change with time as  result of the types changing their properties while still insisting the change isn't big enough to consider the type to become a different type. This view point is used in many models which study the dynamics of the combined population dynamics of types and the slow dynamics of the interactions between types. We already mentioned Jain and Krishna\cite{Jain_Krishna_1998,Jain_Krishna_2002} other examples are the models investigated by Shimada and collaborators\cite{Murase_2010,Shimada_2014} to investigate ecosystem assembly and stability. This approach is related to the broad scope study of "correlated novelties" by Tria et al. \cite{Loreto_2014} in which the expansion of a system into the hitherto unoccupied parts of type space, the adjacent possible, is modelled by use of Polya Urns.

\section{Summary, Conclusion and Out Look}
We used the Tangled Nature framework to emphasised the complexity science aspects of co-evolutionary dynamics. The central point consists in a focus on interdependencies  between the co-evolving components. Whether we consider organisms, bacteria, companies or human agents the behaviour of one agent is always intricately interwoven with the behaviour of some, and often many, of the other surrounding  agents. The generality of the simple dynamical models is related to the fact that the same kind of processes can occur at vastly different levels and in systems that in terms of components (companies versus bacteria) appear to be very different. The focus on the processes  shared between different systems may help us to identify essential mechanism that are not contingent on all the specific details of a system, but may be a consequence of more general and wide spread mechanisms. Think of the normal distribution very often encountered across all sciences. The reason for this is the central limit theorem which identifies some often fulfilled conditions under which sums of numbers properly normalised will be distributed according to the Gaussian functional form. So the normal distribution is not a consequence of some unique properties of a specific system. Complexity science tries to identify similar general principles of broad relevant to systems consisting of many interacting components. The Tangled Nature approach is one example of this research agenda and its relevance to both biological and socio-economical systems as discussed above suggests it is worth to further develop the framework.  Current research attempts to develop TaNa models of evolutionary aspects of tumour growth and of the differences between exogenous and endogenous instabilities in financial systems, just to mention two activities known to the author.

\section{Acknowledgement}
The work on the Tangled Nature framework presented here was very much carried out in collaboration and through discussions with numerous students and colleagues. I am particularly grateful to P. Anderson, E. Arcaute, K. Brinck,  A. Cairoli, K. Christensen, S.A. di Collobiano, J. Grujic, M. Hall, R. Hanel,  D. Jones,  S. Laird, D.J. Lawson,  Y. Murase, D. Piovani, P.A. Rikvold,  R.D. Robliano, T. Shimada, P. Sibani,  S. Thurner and P. Vazquez.

\bibliographystyle{plain}

  
\end{document}